\documentclass{aastex}          
\usepackage{spr-astr-addons}    

\begin{document}

\title{Braneworld cosmology in $f(R,T)$ gravity}

\shorttitle{<Short article title>}
\shortauthors{<Autors et al.>}

\author{P.H.R.S. Moraes$^{1}$\altaffilmark{}} 
\author{R.A.C. Correa$^{2}$\altaffilmark{}}

\affil{$^{1}$ITA - Instituto Tecnol\'ogico de Aeron\'autica - Departamento de F\'isica, 12228-900, S\~ao Jos\'e dos Campos, S\~ao Paulo, Brasil}
\email{moraes.phrs@gmail.com} 
\affil{$^{2}$CCNH, Universidade Federal do ABC, 09210-580, Santo Andr\'{e}, SP,
Brazil}
\email{rafael.couceiro@ufabc.edu.br}

\begin{abstract}
Braneworld scenarios consider our observable universe as a brane embedded in a 5D space, named bulk. In this work, we derive the field equations of a braneworld model in a generalized gravitational theory, namely $f(R,T)$ gravity, with $R$ and $T$ representing the Ricci scalar and the trace of the energy-momentum tensor, respectively. The cosmological parameters obtained from this approach are in agreement with recent constraints from type Ia supernovae data, baryon acoustic oscillations and cosmic microwave background observations, favouring such an alternative description of the universe dynamics.
\end{abstract}


\section{Introduction}\label{sec:int}

General Relativity breaks down at high enough energies. At some time in the early universe, it reached an energy scale which cannot be described by Einstein's theory. At those regimes we need a quantum theory of gravity. Attempts for that are found in the literature as Loop Quantum Gravity (\cite{ashtekar/2004,ashtekar/2011,rovelli/1990,rovelli/1998}) and M-theory (\cite{banks/1997,berman/2011,dasgupta/1999}).

Plenty of efforts have been made in trying to describe the observable universe as a $3-$brane embedded in a $5$D space-time named {\it bulk}. Some of these models rise as a special case of 11-dimensional M-theory. In Randall-Sundrum (RS) I braneworld model (\cite{randall/1999}), such a universe set up, with two branes - the weak brane, where we live in, and the gravity brane - was presented as an optimistic alternative to solve the hierarchy problem. Differently from the other fundamental forces, gravity would not be stuck on the weak brane. It would always ``realize" the effects of an extra compactified dimension. Note that according to Einstein, gravity is connected with the geometry of the (entire) space-time; in this way, it is expected to act in each dimension.

In RSII braneworld model (\cite{randall/1999b}), the weak brane is ``sent to infinite" and we and all the standard model particles are found at the gravity brane. Note that despite gravity ``leaks" through the extra dimension, it is concentrated near the brane, i.e., the graviton probability function has a maximum at the brane and exponentially falls as one moves away from the brane through the fifth dimension, which differently from RSI case, can be large. Indeed, this happens because the extra dimension in RSII model is considered to be warped. Moreover, RSII braneworld is capable of recovering, in a certain energy regime, the newtonian gravitational potential, as carefully discussed in (\cite{randall/1999b}).

A lot of cosmological models have been derived from RSII model (check, for instance, (\cite{binetruy/2000,flanagan/2000,kim/2000,apostolopoulos/2006,campos/2003}) and references therein, and also (\cite{brax/2004}) for a seminal braneworld cosmology review).

RSII model has also been considered as the universe set up for generalized models of gravity, such as $f(R)$ theories (\cite{nojiri/2003,shamir/2010,masoudi/2013,hussain/2011}), as one can see, for instance, in (\cite{balcerzak/2011,bazeia/2013,xu/2015}).

Recently, a more general theory of gravity was proposed by T. Harko and collaborators, for which, besides the general dependence on the Ricci scalar $R$ as in $f(R)$ theories, the gravitational part of the action also depends on a generic function of $T$, the trace of the energy-momentum tensor, namely the $f(R,T)$ gravity (\cite{harko/2011}). Although plenty of extradimensional cosmological models have been derived from $f(R,T)$ theories (e.g. (\cite{moraes/2015,moraes/2014,ram/2013,reddy/2013,rao/2015,reddy/2012,samanta/2013})), none of them has considered braneworld scenarios so far. Therefore, it seems reasonable and promising to consider $f(R,T)$ gravity in the scope of braneworld scenarios, more specifically, the RSII model, which is the aim of the present work.

In fact, $f(R,T)$ gravity has already been considered in braneworld models (\cite{bazeia/2015,cm/2015}). However, such applications were made in thick (non-RS) brane models and no cosmological scenarios have been derived.

Note that the $f(R,T)$ theories predict a coupling between matter, through the dependence on $T$, and geometry, through the dependence on $R$. The $T$-dependence could be interpreted as inducted from the existence of quantum effects (\cite{harko/2011}). In fact, this dependence links with known illustrious proposals such as geometrical curvature inducing matter and geometrical origin of matter content in the universe (\cite{shabani/2013,farhoudi/2005}). Meanwhile, the RSII bulk contains only gravity (geometry), and the brane (matter) arises as a geometrical manifestation of such an empty 5D space-time. The match between these features justify and make promising the consideration of $f(R,T)$ theories in braneworlds.

As it will be shown below, the braneworld approach raises some new quantities, not found in standard cosmology, like the 5D  bulk cosmological constant $\Lambda^{(5)}$ and the brane tension $\sigma$. In (\cite{mm/2014}) it was presented a pioneer form for constraining the values that the brane tension cosmological parameter can assume, from the study of the sign of gravitational waves emitted by a binary system of neutron stars.

It is worth mentioning that recently, in the literature, it has been common to see efforts focused on probing the existence of a braneworld extra dimension. Since the photons are confined to the brane, the extra dimension cannot be observed and its probing is made indirectly, as one can check in (\cite{simonetti/2011,yagi/2011,johannsen/2009,farajollah/2013,rahimov/2011}), for instance.

\section{The $f(R,T)$ gravity}\label{sec:frt}

Recently proposed by T. Harko et al., the $f(R,T)$ gravity considers the gravitational part of the action as being dependent on a function of $R$, the Ricci scalar, and $T$, the trace of the energy-momentum tensor $T_{\mu\nu}$ (\cite{harko/2011}). The dependence on $T$ is justified by the consideration of quantum effects. In this way, the gravitational part of the action reads

\begin{equation}\label{frt1}
S_G=\frac{1}{16\pi}\int d^4x\sqrt{-g}f(R,T),
\end{equation}
with $f(R,T)$ being the function of $R$ and $T$, $g$ the determinant of the metric and we will work with units such that $c=G=1$.

By varying Eq.(\ref{frt1}) with respect to the metric, one obtains the following field equations (FEs):

\begin{align}\label{frt2}
&f_R(R,T)R_{\mu\nu}-\frac{1}{2}f(R,T)g_{\mu\nu}+(g_{\mu\nu}\Box-\nabla_\mu\nabla_\nu)\nonumber \\
&f_R(R,T)=8\pi T_{\mu\nu}-f_T(R,T)T_{\mu\nu}-f_T(R,T)\Theta_{\mu\nu}.
\end{align}
In (\ref{frt2}), as usually, $R_{\mu\nu}$ stands for the Ricci tensor, $T_{\mu\nu}=g_{\mu\nu}\mathcal{L}_m-2\partial\mathcal{L}_m/\partial g^{\mu\nu}$ is the energy-momentum tensor, with $\mathcal{L}_m$ being the matter lagrangian, $\nabla_\mu$ is the covariant derivative with respect to the symmetric connection associated to $g_{\mu\nu}$ with $\mu,\nu$ running from $0$ to $3$, as originally proposed by the authors, $f_R(R,T)\equiv\partial f(R,T)/\partial R$, $f_T(R,T)\equiv\partial f(R,T)/\partial T$, $\Box\equiv\partial_\mu(\sqrt{-g}g^{\mu\nu}\partial_\nu)/\sqrt{-g}$ and $\Theta_{\mu\nu}\equiv g^{\alpha\beta}\delta T_{\alpha\beta}/\delta g^{\mu\nu}$. 

It can be seen, from several cosmological models obtained via $f(R,T)$ gravity, such as (\cite{rao/2015,baffou/2015,shabani/2014,moraes/2015b}), that the extra terms\footnote{``Extra terms" when compared to standard gravity field equations.} in the FEs (\ref{frt2}) may be the responsible for the present cosmic acceleration the universe is passing through (\cite{riess/1998,perlmutter/1999}). Such a cosmological feature in standard cosmology is justified by the existence of the cosmological constant $\Lambda$, which yields a sort of ``anti"-gravitational effect on the Einstein's FEs. In this way, $f(R,T)$ gravity is able to predict the cosmic acceleration without the necessity of invoking the cosmological constant. Consequently, such a scenario is free of the cosmological constant problem (\cite{weinberg/1989,peebles/2003,luongo/2012}).

\section{The field equations of the $f(R,T)$ braneworld model}\label{sec:fe}

The total action in RSII model consists of the bulk and brane actions, as (\cite{brax/2004})

\begin{equation}\label{rs1}
S_{bulk}=-\int d^{5}x\sqrt{-g^{(5)}}\left[\frac{R^{(5)}}{16\pi}+\Lambda^{(5)}\right],
\end{equation}
\begin{equation}\label{rs2}
S_{brane}=\int d^{4}x\sqrt{-g}(-\sigma).
\end{equation}
In the equations above, $g^{(5)}$ is the determinant of the 5D metric, $R^{(5)}$ is the 5D Ricci scalar, $\Lambda^{(5)}$ is the bulk cosmological constant, $g$ the determinant of the 4D metric and $\sigma$ is the brane tension.

Let me assume, as in (\cite{brax/2004}), that the RSII bulk metric reads 

\begin{equation}\label{m1}
ds^{2}=a^{2}(t,y)b^{2}(t,y)(dt^{2}-dy^{2})-a^{2}(t,y)\eta_{ij}dx^{i}dx^{j}.
\end{equation}
In (\ref{m1}), $a$ and $b$ are the scale factors and $\eta_{ij}$ is the Minkowski metric with $i,j=1,2,3$, so that (\ref{m1}) is consistent with a homogeneous and isotropic brane located at $y=0$, with $y$ being the extra space-like dimension.

The non-null components of the Einstein tensor in the bulk for (\ref{m1}) are 

\begin{equation}\label{m2}
G_0^{0}=3\left[2\left(\frac{\dot{a}}{a}\right)^{2}+\frac{\dot{a}}{a}\frac{\dot{b}}{b}-\frac{a''}{a}+\frac{a'}{a}\frac{b'}{b}+\tilde{\sigma} b^{2}\right],
\end{equation}
\begin{equation}\label{m3}
G_1^{1}=3\frac{\ddot{a}}{a}+\frac{\ddot{b}}{b}-\left(\frac{\dot{b}}{b}\right)^{2}-3\frac{a''}{a}-\frac{b''}{b}+\left(\frac{b'}{b}\right)^{2}+\tilde{\sigma} b^{2},
\end{equation}
\begin{equation}\label{m4}
G_2^{2}=G_3^{3}=G_1^{1},
\end{equation}
\begin{equation}\label{m5}
G_0^{4}=3\left(-\frac{\dot{a}'}{a}+2\frac{\dot{a}a'}{a^{2}}+\frac{\dot{a}}{a}\frac{b'}{b}+\frac{a'}{a}\frac{\dot{b}}{b}\right),
\end{equation}
\begin{equation}\label{m6}
G_4^{4}=3\left[\frac{\ddot{a}}{a}-\frac{\dot{a}}{a}\frac{\dot{b}}{b}-2\left(\frac{a'}{a}\right)^{2}-\frac{a'}{a}\frac{b'}{b}+\tilde{\sigma} b^{2}\right],
\end{equation}
with dots and primes standing for derivations with respect to time and extra dimension, respectively, and $\tilde{\sigma}\equiv 8\pi\sigma/\sqrt{6}$.

Naturally, those non-null components of the Einstein tensor must be related to the matter-energy content of the universe. In order to construct such relations, let me develop Eq.(\ref{frt2}) in $5$D, by assuming $f(R,T)=R+2\lambda T$, with $\lambda$ a constant. Such an $f(R,T)$ functional form has been widely explored in cosmology, as it can be seen, for instance, in (\cite{harko/2011,moraes/2015,moraes/2015b,moraes/2014,kumar/2015,mahanta/2014}). In 5D, one has, from (\ref{frt2}):

\begin{equation}\label{frtx}
G_{AB}=8\pi T_{AB}+\lambda Tg_{AB}+2\lambda(T_{AB}+Pg_{AB}),
\end{equation}
for which $A,B$ run from $0$ to $4$ and $P$ is the pressure of the universe.

It should be stressed that for the energy-momentum tensor of a perfect fluid on the rhs of Eq.(\ref{frtx}), one must consider the contributions from both the bulk and the brane. For instance, in the above equation, $P=p_{bulk}+p$, with $p$ being the brane pressure. In the same way, below, $\rho$ will stand for the brane density. Therefore, with the help of Eqs.(\ref{m2})-(\ref{m6}), the development of (\ref{frtx}) yields

\begin{align}\label{m7}
&3\left[2\left(\frac{\dot{a}}{a}\right)^{2}+\frac{\dot{a}}{a}\frac{\dot{b}}{b}-\frac{a''}{a}+\frac{a'}{a}\frac{b'}{b}+ \tilde{\sigma} b^{2}\right]=4(2\pi+\lambda)\Lambda^{(5)}\nonumber \\
&+(8\pi+3\lambda)\rho-\lambda p,
\end{align}
\begin{align}\label{m8}
&3\frac{\ddot{a}}{a}+\frac{\ddot{b}}{b}-\left(\frac{\dot{b}}{b}\right)^{2}-3\frac{a''}{a}-\frac{b''}{b}+\left(\frac{b'}{b}\right)^{2}+\tilde{\sigma} b^{2}=8\pi\Lambda^{(5)}\nonumber\\
&+\lambda\rho-(8\pi+\lambda)p,
\end{align}
along with the following constraint relations

\begin{equation}\label{m9}
\dot{a}'=2\dot{a}a'+\dot{a}+a'\frac{b'}{b},
\end{equation}
\begin{equation}\label{m10}
\sigma=\frac{\sqrt{6}}{8\pi b^{2}}\left[-\frac{\ddot{a}}{a}+\frac{\dot{a}}{a}\frac{\dot{b}}{b}+2\left(\frac{a'}{a}\right)^{2}+\frac{a'}{a}\frac{b'}{b}\right].
\end{equation}
It should be noted that the equations above stand for the brane location. Moreover, I have used the relation $\rho_{bulk}=-p_{bulk}=\Lambda^{(5)}$, which is assumed in RS models (\cite{brax/2004,binetruy/2000}), and both $T_4^{0}$ and $T_4^{4}$ are null, the latter because, as mentioned in Section \ref{sec:int}, the extra dimension contains only gravity. 

\section{A matter-dominated universe in $f(R,T)$ braneworld scenario}\label{sec:mdu}

When deriving the cosmological solutions for the $f(R,T)$ brane model, I will assume the universe is dominated by matter, with equation of state (EoS) $p=0$ to be substituted in (\ref{m7})-(\ref{m8}). Such a consideration along with Eq.(\ref{m10}) make us able to write

\begin{equation}\label{m14}
\frac{5}{2}\left(\frac{\ddot{a}}{a}-\frac{a''}{a}\right)-\left(\frac{\dot{a}}{a}\right)^{2}+\left(\frac{a'}{a}\right)^{2}=-\tilde{\Lambda}^{(5)},
\end{equation} 
with $\tilde{\Lambda}^{(5)}\equiv(-44\pi/9)\Lambda^{(5)}$ and $\lambda=8\pi/3$. In (\ref{m14}), as in (\cite{binetruy/2000}),  the fifth dimension was considered to be static.

In order to solve Eq.(\ref{m14}) we will use the method of separation of variables. By taking $a(t,y)=\theta(t)\varepsilon(y)$, with $\theta(t)$ and $\varepsilon(y)$ being functions of $t$ and $y$ only, respectively, we obtain from (\ref{m14}):

\begin{equation}\label{m15}
\frac{5}{2}\frac{\ddot{\theta}}{\theta}-\left(\frac{\dot{\theta}}{\theta}\right)^{2}=0,
\end{equation}
\begin{equation}\label{m16}
\frac{5}{2}\frac{\varepsilon''}{\varepsilon}-\left(\frac{\varepsilon'}{\varepsilon}\right)^{2}=\tilde{\Lambda}^{(5)},
\end{equation}
for which we have taken the constant of separation to be $-\tilde{\Lambda}^{(5)}$.

The solution of Eq.(\ref{m15}) is

\begin{equation}\label{m17}
\theta (t)=C_{1}(3t+{C}_{2})^{5/3},
\end{equation}
with $C_1$ and $C_2$ being constants. 

On the other hand, the solution of (\ref{m16}) will depend on the sign of the constant $\tilde{\Lambda}^{(5)}$. By choosing positive values for $\tilde{\Lambda}^{(5)}$, the solutions will be exponentials of $y$, while when $\tilde{\Lambda}^{(5)}$ is negative, an oscillatory dependence for $y$ in the scale factor is obtained as $\varepsilon\sim\cos y$. An oscillatory dependence for such a scale factor is non-physical, so we shall discard the case $\tilde{\Lambda}^{(5)}<0$.

If we choose to work with the gauge $\tilde{\Lambda}^{(5)}=1$, we have

\begin{equation}\label{m18}
\varepsilon (y)=C_3(C_4e^{\alpha y}+1)^{5}e^{-\beta y},
\end{equation}
with $C_3$ and $C_4$ being constants, $\alpha =2\sqrt{6}/5$ and $\beta =\sqrt{6}/3$. 

Now, from Eqs.(\ref{m17})-(\ref{m18}), we can write the scalar factor $a(t,y)$ as

\begin{equation}
a(t,y)=C_0(3t+C_2)^{5/3}(C_4e^{\alpha y}+1)^{5}e^{-\beta y},  \label{ad3}
\end{equation}
with $C_0\equiv C_1C_3$. Here, it is important to remark that the solutions of Eqs.(\ref{m15}) and (\ref{m16}), given by Eqs.(\ref{m17}) and (\ref{m18}), respectively, contain four arbitrary integration constants, which are determined by initial conditions. By taking $a(0,y)=0$ yields $C_0\neq 0$ and $C_2=0$. Moreover, as we are interested in determining the Hubble
factor $H=\dot{a}/a$, it is not necessary to find the values of $C_0$ and 
$C_4$, since such a ratio will remove the dependence on these constants.

From Eq.(\ref{ad3}), one is able to plot the Hubble parameter $H=\dot{a}/a$ of the $f(R,T)$ braneworld model. Such a cosmological parameter is depicted below.

\begin{figure}[h!]
\vspace{1cm}
\includegraphics[{height=05cm,angle=00}]{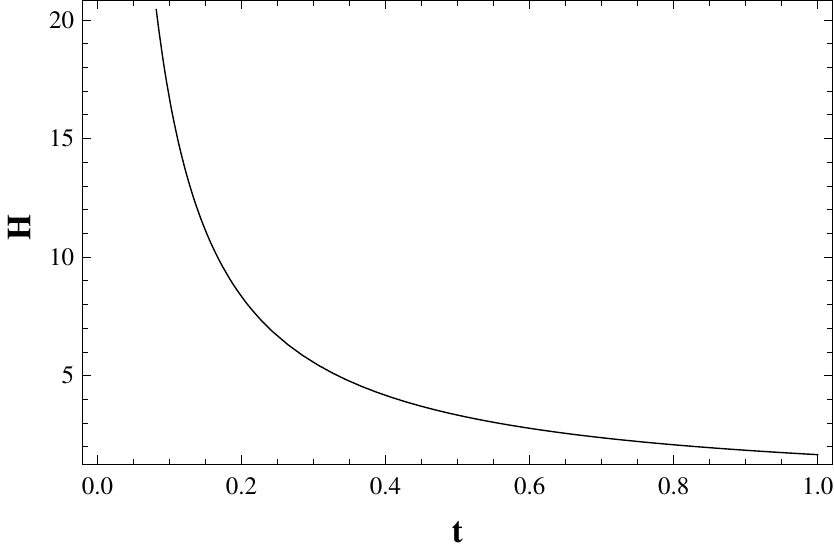}
\vspace{0.3cm}
\caption{Plot of the Hubble parameter.}
\label{FIG1}
\end{figure} 

Furthermore, in possession of Eq.(\ref{ad3}), one is also able to calculate the deceleration parameter $q=-a\ddot{a}/\dot{a}^{2}$, so that the universe expansion is accelerating when $q<0$ and decelerating when $q>0$. From Eq.(\ref{ad3}), one obtains $q=-0.4$, in accordance with the acceleration of the universe expansion.

\section{Conclusions}\label{sec:dis}

In this work, we have presented a novel cosmological scenario, which is derived from the consideration of the $f(R,T)$ theory of gravity in the RSII braneworld. Although plenty of extradimensional cosmological models have been derived in $f(R,T)$ gravity (check, for instance, (\cite{moraes/2015,moraes/2014,ram/2013})), those were not obtained from the braneworld scenario consideration. Rather, they were obtained from Kaluza-Klein models (\cite{overduin/1997,moraes/2016}). Although the unification of $f(R,T)$ gravity with braneworld scenarios was made (\cite{bazeia/2015,cm/2015}), in such works, the braneworld set up was not RS like, and last, but not least, no cosmological solutions were obtained.

In Section \ref{sec:mdu}, the solutions of Eq.(\ref{m14}) favoured a negative bulk cosmological constant $\Lambda^{(5)}$ (remind that $\Lambda^{(5)}>0$ yields a non-physical scale factor). It is well known that the bulk cosmological constant must, indeed, be negative (\cite{brax/2004,randall/1999,randall/1999b,campos/2003,germani/2001}). In fact, a positive bulk cosmological constant would accelerate the ``leaking" of gravity through the extra dimension, making it harder to locate such a force on the brane. While a negative bulk cosmological constant is sometimes imposed to braneworld cosmological models, in the present work, this feature has emerged naturally.

Still in Section \ref{sec:mdu} we have derived the resultant cosmological parameters of the model. Fig.\ref{FIG1} reveals a well-behaved evolution for the Hubble parameter. Firstly note that, from standard cosmology, $H\sim t_H^{-1}$, with $t_H$ being the Hubble time. It can be seen from Fig.\ref{FIG1} that such a feature is conserved in the $f(R,T)$ brane model. Furthermore, $H$ is restricted to positive values, assuring the universe expansion.

The deceleration parameter is defined in such a way that if $q<0$, the universe expansion is accelerating. Type Ia Supernovae (\cite{riess/1998,perlmutter/1999}) and temperature fluctuations on the cosmic microwave background radiation observations (\cite{hinshaw/2013}) confirm our universe is undergoing a phase of accelerated expansion. In Section \ref{sec:mdu}, the result predicted by the model for the present value of the deceleration parameter was $q=-0.4$, in accordance with an accelerated expansion. 

Moreover, $q=-0.4$ is in the range of accepted values for $q$ at present as one can see in (\cite{hinshaw/2013,giostri/2012}), in which for the latter, the authors have combined Baryon Acoustic Oscillations observations with Supernovae Ia data to constrain the values of the deceleration parameter. Here, it is worth mentioning that the values of the constants $C_0$ and $C_4$ have no effect on the numerical value of $q$ in the present model. In fact, there was no necessity of assuming any values for them.

A value for $q$ which is also in accordance with (\cite{giostri/2012}) was already found by considering $f(R,T)$ gravity in another extradimensional model, the Kaluza-Klein gravity (\cite{overduin/1997}). In that approach, in order to obtain a negative accepted value for $q$ it was necessary to assume the EoS $p=-\rho$ in the model FEs (\cite{moraes/2014}). Such an EoS is in agreement with recent Wilkinson Mapping Anisotropy Probe observations, which predict $\omega=-1.073^{+0.090}_{-0.089}$ at present for a cosmological constant dominated universe (\cite{hinshaw/2013}). On the other hand, in the present $f(R,T)$ braneworld model, it was not necessary to assume the EoS above. Recall that in Section \ref{sec:mdu}, I have assumed $p=0$. Such an EoS stands for a matter dominated universe, i.e., there was no need of assuming any kind of exotic fluid for the universe composition and nevertheless, an accelerated expansion was obtained.

Furthermore, it is worth stressing that some extradimensional $f(R,T)$ models found in the literature could not be able to generate a negative deceleration parameter, predicting, this way, a decelerating expansion (\cite{reddy/2012,reddy/2013}). On the other hand, from a generic functional form for $f(R,T)$ and without the necessity of invoking an exotic EoS for the universe, the present model was able to predict the cosmic acceleration.

\acknowledgments

PHRSM would like to thank S\~ao Paulo Research Foundation (FAPESP), grant 2015/08476-0, for financial support. PHRSM is also thankful to the anonymous referee, for his/her comments, which made the mathematics of the paper more robust, yielding a more realistic cosmological model. RACC thanks to UFABC and CAPES for financial support

\bibliographystyle{spr-mp-nameyear-cnd}  

\end{document}